\documentclass[%
 aip,
 jcp,
 amsmath,amssymb,
 reprint,%
]{revtex4-1}

\usepackage{braket}
\usepackage{graphicx}
\usepackage{hyperref}
\usepackage{bm} 

\usepackage[utf8]{inputenc}
\usepackage[T1]{fontenc}
\usepackage{mathptmx}

\begin{document}

\title{Plane-wave approach to the exact van der Waals interaction between colloid particles}

\author{Benjamin Spreng}
\affiliation{Universität Augsburg, Institut für Physik, 86135 Augsburg, Germany}
\author{Paulo A. Maia Neto}
\affiliation{Instituto de Física, Universidade Federal do Rio de Janeiro, CP 68528, Rio de Janeiro RJ 21941-909, Brazil}
\author{Gert-Ludwig Ingold}
\affiliation{Universität Augsburg, Institut für Physik, 86135 Augsburg, Germany}

\date{\today}

\begin{abstract}
The numerically exact evaluation of the van der Waals interaction,
also known as Casimir interaction when
including retardation effects, constitutes a challenging task. We
present a new approach based on the plane-wave basis and demonstrate
that it possesses advantages over the more commonly used multipole
basis. The rotational symmetry of the plane-sphere and sphere-sphere
geometries can be exploited by means of a discrete Fourier transform.
The new technique is applied to a study of the interaction between a
colloid particle made of polystyrene or mercury and another polystyrene
sphere or a polystyrene wall in an aqueous solution.  Special attention
is paid to the influence of screening caused by a variable salt
concentration in the medium. It is found that in particular for low
salt concentrations the error implied by the proximity force
approximation is larger than usually assumed. For a mercury droplet,
a repulsive interaction is found for sufficiently large distances provided
screening is negligible. We emphasize that the effective Hamaker parameter
depends significantly on the scattering geometry on which it is based.
\end{abstract}

\maketitle

\section{Introduction}

Aqueous colloidal suspensions play an important role in our daily lives, both
in natural substances as well as in manifold industrial applications. For an
understanding of their physical properties as well as for the design of such
colloidal systems, a theoretical description of the relevant forces is
required.  In aqueous suspensions repulsive double layer electrostatic forces play a major
role. \cite{Butt2010, Israelachvili2011}
An important omnipresent interaction for all
colloidal systems is the van der Waals force, \cite{Israelachvili2011} which
is typically attractive but can be repulsive\cite{Dzyaloshinskii1961, Milling1996, Munday2009,
Tabor2011, Thiyam2018, Esteso2019} for certain combinations of materials. These two forces are essential
for understanding the stability of colloids.

Colloidal suspensions are composed of nano- to micrometer-size objects. For
colloid particles less than $10\,\text{nm}$ apart, the fluctuating
electromagnetic interaction can typically be treated as instantaneous while for
larger separations retardation effects need to be taken into account.
\cite{Butt2010} The retarded van der Waals force is often also referred to as Casimir
force.\cite{Casimir1948, Lifshitz1956, Bordag2009} In this paper, we present
exact theoretical results for the van der Waals force covering the entire
distance range of experimental interest, while focusing on larger distances. In
addition to retardation, non-trivial geometry effects become more important as
the distance increases, making commonly employed approximations increasingly
inaccurate.

As a minimal model for the study of aqueous colloidal systems, the interaction
of two spherical particles in a solvent or the interaction between a spherical
particle and a wall can be considered. Such situations have been realized in
numerous experiments. \cite{Bevan1999, Hansen2005, Wodka2014, Ether2015, Montes2017}
The plane-sphere and sphere-sphere setup are also typical for experiments
exploring Casimir forces across vacuum or air. \cite{Klimchitskaya2009, Decca2011, Lamoreaux2011, Klimchitskaya2020}

The finite curvature of the spherical particles is usually accounted for by
means of the proximity force approximation (PFA), also known as Derjaguin
approximation. \cite{Derjaguin1934} Within the PFA, the van der Waals interaction energy is
calculated by averaging the energies of parallel planes over the local distances.
\cite{Parsegian2005} The PFA can be understood as an asymptotic result where the
sphere radius represents the largest length scale. \cite{Spreng2018} In order
to determine its range of validity one needs to determine subleading correction
terms or make use of numerical techniques.

Analytically, corrections to the PFA can be found through asymptotic expansions
of the scattering-theoretical expression for the van der Waals energy
\cite{Bordag2008, Teo2011, Teo2012, Teo2013,Henning2019} or by means of a derivative
expansion.\cite{Fosco2011, BimonteEPL2012, BimonteAPL2012, Fosco2014,
Fosco2015} Also, more recently a semi-analytical method utilizing the
derivative expansion has been proposed.\cite{Bimonte2018a, Bimonte2018b}

Numerical methods for computing the van der Waals interaction complement analytical
results beyond the PFA. They do not only serve as a quality check for
approximations, but can yield exact results valid for any separation between
the objects. Numerical approaches applicable to general geometries include the
boundary-element method \cite{Reid2015} and simulations based on the
finite-difference time-domain method. \cite{Oskooi2010}

Specializing on spheres and possibly planes, larger aspect ratios between
sphere radius and surface-to-surface distance are accessible by making use of the
scattering formalism. \cite{Lambrecht2006, Emig2007}  For such geometries,
bispherical coordinates might appear as an efficient tool to derive the
relevant scattering operators. However, only the Laplace equation and not the
Helmholtz equation is separable in these coordinates.\cite{Boyer76} In
practice, their use is thus limited to the zero-frequency contribution of the
van der Waals interaction. \cite{Bimonte2017, Bimonte2018b}  The contribution
of arbitrary frequencies is computed by expressing the scattering operators
referring to the individual surfaces in appropriately chosen local coordinate
systems.  For numerical purposes, the scattering operators as well as the
translation operators connecting the two coordinate systems have to be
expressed in a suitable basis.  Spherical multipoles have been used in numerous
studies of scattering geometries composed of a plane and a sphere,
\cite{MaiaNeto2008,Emig2008,Canaguier-Durand10,
Canaguier-Durand-thesis,Canaguier-Durand12} two spheres
\cite{Emig2007,Umrath2015} or a grating and a sphere. \cite{Messina2015} While
for large distances, it is sufficient to take into account only a few
multipoles, the situation changes dramatically when experimentally relevant
aspect ratios are considered. This range became accessible only recently by
symmetrization of the scattering operator and employing hierarchical low-rank
approximation techniques. \cite{Hartmann2017, Hartmann2018, HartmannThesis}

In this paper, we will develop an alternative numerical approach making use of
the plane-wave basis, which is better adapted to capture the effect of near
specular reflection in the vicinity of the WKB scattering regime,
\cite{Spreng2018} while still allowing for arbitrary temperatures and
materials. The convergence properties of our approach will turn out to be far
superior to those found for the spherical multipole basis.  Often, scattering
operators are already known in the plane-wave basis as scattering
amplitudes~\cite{vdHulst} derived for different geometries by making use of the
appropriate coordinate systems.\cite{Boyer76}  In addition, this basis makes
translations between  the coordinate systems employed for each particle
particularly simple. In comparison with the spherical multipole basis, the
non-discreteness of the plane-wave basis might appear as an important drawback.
However, as we will discuss in this paper, the problem can be circumvented in a
natural way by  means of a Nystr\"om discretization of the plane-wave momenta.
As a consequence, our method based on the
the plane-wave basis turns out to be easier to implement than the more standard multipolar approach.

As an illustration of this new numerical plane-wave approach, we will consider two
polystyrene microspheres in an aqueous solution and compare the PFA with the
numerically computed van der Waals interaction. In contrast to the work by Thennadil
and Garcia-Rubio, \cite{Thennadil2001} our comparison is exact and goes beyond
the perturbative approach developed by Langbein. \cite{Langbein1970,
Langbein1974}  Likewise, we study the van der Waals interaction between a
polystyrene microsphere and a polystyrene wall.  The exact results for the two
polystyrene surfaces in the plane-sphere and sphere-sphere geometry
then allow us to study the geometry dependence of the Hamaker parameter.  We
finally conduct an analogous analysis for the system of a mercury microsphere
and a polystyrene surface.  The van der Waals interaction of such systems can be
repulsive or attractive depending on how the distance between the surfaces
compares with the Debye screening length.\cite{Ether2015}

This paper is structured as follows. In Sec.~\ref{sec:2}, the numerical method
is introduced first for arbitrary scatterers and then specialized to geometries
with a cylindrical symmetry. In Sec.~\ref{sec:3}, we apply the method to
the plane-sphere and sphere-sphere geometries. The convergence properties of
our numerical approach are studied for both examples. Furthermore, for the
plane-sphere geometry we compare the computational time needed with the
plane-wave approach and a reference implementation based on spherical
multipoles. \cite{HartmannJOSS} In Sec.~\ref{sec:4}, we apply the plane-wave
method to the analysis of the van der Waals interaction in colloidal systems containing
polystyrene or mercury spheres. Sec.~\ref{sec:5} contains concluding remarks
and the appendices provide technical details supporting the main text of the
paper.

\section{van der Waals interaction within the plane-wave basis}
\label{sec:2}
\subsection{Geometry of two arbitrary objects}

\begin{figure}
 \includegraphics[width=0.4\columnwidth]{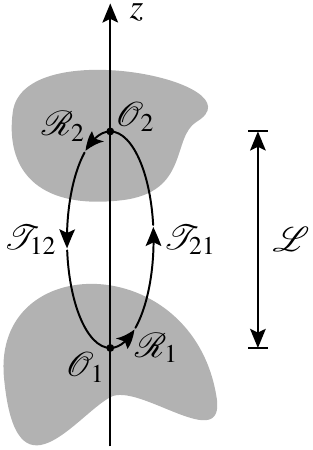}
 \caption{Schematic representation of the scattering geometry containing two objects 1 and 2.
	  Their respective reference points $\mathcal{O}_1$ and $\mathcal{O}_2$ define the $z$-axis
	  and are separated by a distance $\mathcal{L}$.}
 \label{fig:potato_geometry}
\end{figure}

We will consider two arbitrary objects 1 and 2 immersed in a homogeneous medium
at temperature $T$ and assume that the objects can be spatially separated by a
plane. Within the scattering-theoretical approach, the van der Waals free
energy of this setup is given by~\cite{Lambrecht2006}
\begin{equation}\label{eq:F}
\mathcal{F} = \frac{k_B T}{2} \sum_{n=-\infty}^\infty
              \log\det\left[1-\mathcal{M}(\vert\xi_n\vert)\right]
\end{equation}
with the Matsubara frequencies $\xi_n=2\pi n k_B
T/\hbar$ and the round-trip operator
\begin{equation}\label{eq:roundtrip_operator}
\mathcal{M} =  \mathcal{T}_{12}\mathcal{R}_2 \mathcal{T}_{21}\mathcal{R}_1\,.
\end{equation}
$\mathcal{R}_j$ denotes the reflection operator of object $j$ with respect to
the reference point $\mathcal{O}_j$ while $\mathcal{T}_{21}$ carries out
the translation from the reference point of object $1$ to the one of object
$2$, and vice versa for $\mathcal{T}_{12}$. The two reference points
$\mathcal{O}_1$ and $\mathcal{O}_2$ are separated by a distance $\mathcal{L}$
and define the $z$-axis of our coordinate system.
Figure~\ref{fig:potato_geometry} schematically depicts the full round trip
described by $\mathcal{M}$.

By taking the derivative of \eqref{eq:F} with respect to the distance
$\mathcal{L}$, one obtains the van der Waals force
\begin{equation}
F = -\partial_\mathcal{L}\mathcal{F}
  = \frac{k_B T}{2} \sum_{n=-\infty}^\infty \mathrm{tr}\left[
	  \frac{\partial_\mathcal{L}\mathcal{M}}{1-\mathcal{M}}\right]
\end{equation}
where we dropped the dependence of the round-trip operator on the Matsubara
frequencies $\xi_n$. A corresponding formula for the force gradient can be found for
instance in Ref.~\onlinecite{Bimonte2018b}.

The frequency of a plane wave is conserved during a round trip between the two
objects.  Thus, it is convenient to employ the so-called angular spectral
representation.\cite{NietoVesperinas} Within this representation, $\{\ket{\xi,
\mathbf{k}, \phi, p}\}$ denotes the plane-wave basis at imaginary frequency
$\xi$ where $\mathbf{k}$ is the transverse wave vector perpendicular to the
$z$-axis, $\phi=\pm 1$ the upward or downward propagation direction and
$p=\mathrm{TE},\mathrm{TM}$ the polarization.  Because $\xi$ remains unchanged
during the round trip and $\phi$ only changes sign after reflection off an
object, we shorten the notation of the basis elements to $\ket{\mathbf{k},p}$.

The translation operators $\mathcal{T}_{12}$ and $\mathcal{T}_{21}$ are
diagonal in the plane-wave basis with matrix elements $e^{-\kappa\mathcal{L}}$.
The axial wave vector after Wick rotation is given by
\begin{equation}
\label{eq:kappa}
 \kappa=\left(\epsilon_m\frac{\xi^2}{c^2}+\mathbf{k}^2\right)^{1/2}\,,
\end{equation}
where $\epsilon_m$ denotes the relative permittivity of the medium between the
two objects. In contrast, the reflection operators $\mathcal{R}_1$ and
$\mathcal{R}_2$ and thus the round-trip operator $\mathcal{M}$ will not be
diagonal in general. The latter operators are integral operators which can be
expressed in terms of their respective kernel functions.  For example the
round-trip operator can be written as
\begin{equation}\label{eq:integraloperator}
 \mathcal{M}\ket{\mathbf{k}, p} = \sum_{p'} \int \frac{d\mathbf{k}'}{(2\pi)^2}\,
	K_\mathcal{M}(\mathbf{k'}, p'; \mathbf{k}, p) \ket{\mathbf{k}',p'}
\end{equation}
with its kernel function
\begin{multline}\label{eq:general-kernel}
K_\mathcal{M}(\mathbf{k'}, p'; \mathbf{k}, p) = \\
e^{-\kappa'\mathcal{L}}\sum_{p''} \int \frac{d\mathbf{k}''}{(2\pi)^2}\,K_{\mathcal{R}_2}(\mathbf{k'}, p'; \mathbf{k''}, p'')
e^{-\kappa''\mathcal{L}}\\
\times K_{\mathcal{R}_1}(\mathbf{k''}, p''; \mathbf{k}, p)\,,
\end{multline}
where $K_{\mathcal{R}_j}$ is the kernel of the reflection operator
$\mathcal{R}_j$ for $j=1,2$ and $\kappa'$ as well as $\kappa''$ are
defined according to (\ref{eq:kappa}). Note that the kernel functions depend
also on the frequency $\xi$ which is suppressed in the arguments to not overload
the notation.

For numerical purposes, a Nystr\"om discretization needs to be applied to
the integral appearing in \eqref{eq:integraloperator}. The action of the
round-trip operator is then expressed in terms of a finite matrix
\begin{equation}\label{eq:integraloperatordiscretized}
\mathcal{M}\ket{\mathbf{k}_\alpha, p} =
 \sum_{p',\alpha'} \frac{w_{\alpha'}}{(2\pi)^2} K_\mathcal{M}(\mathbf{k}_{\alpha'}, p'; \mathbf{k}_\alpha, p)
 \ket{\mathbf{k}_{\alpha'},p'}
\end{equation}
with the nodes $\mathbf{k}_\alpha$ and weights $w_\alpha$ of a quadrature rule
for the two-dimensional integral. The indices $\alpha$ and $\alpha'$ represent
tuples $(i,j)$ where $i$ and $j$ are indices from the quadrature rules of
the corresponding one-dimensional integrals.

Within this approximation the matrix elements of the round-trip operator become
the corresponding kernel function multiplied by the quadrature weights
\cite{Bornemann2010}
\begin{equation}
\braket{\mathbf{k}_{\alpha'}, p' \vert \mathcal{M} \vert \mathbf{k}_\alpha, p} =
 \frac{w_{\alpha'}}{(2\pi)^2} K_\mathcal{M}(\mathbf{k}_{\alpha'}, p'; \mathbf{k}_\alpha, p)\,.
\end{equation}
After discretization the matrix elements thus form a three-fold block matrix
with respect to the two indices constituting the tuple $\alpha=(i,j)$ and the
polarization $p$.

In general, the reflection operators pertaining to the two objects are
non-diagonal in the plane-wave basis as is the case for example in the geometry
of two spheres. The integral over $\mathbf{k}''$ appearing inside the kernel of
the round-trip operator~\eqref{eq:general-kernel} can usually not be performed
analytically. Thus, for numerical applications this integral needs to be
discretized as well where the quadrature rule may differ from the one chosen in
\eqref{eq:integraloperatordiscretized}. The round-trip matrix can then be
expressed in terms of a product of two block matrices representing the
reflection operators.

\subsection{Geometry with cylindrical symmetry}
\label{sec:cylindrical_symmetry}

Casimir and van der Waals experiments are often carried out in set-ups with a
certain symmetry.  The cylindrical symmetry is particularly common as it
appears in the plane-sphere and sphere-sphere geometries. At first sight, the
spherical-wave basis appears to be better adapted for those geometries since
the angular momentum index $m$ is conserved through the round trip, yielding a
block-diagonal round-trip matrix. However, this symmetry can also be exploited
in the plane-wave basis as will be explained in the following.

For a cylindrically symmetric geometry it is natural to express the transverse
wave vector $\mathbf{k}$ in polar coordinates with radial component $k$ and
angular component $\varphi$, where the latter is relevant for the following
considerations. A suitable quadrature rule for the integration over $\varphi$
is the trapezoidal rule which at order $M$ has nodes at $\varphi_j= (2\pi/M)j$
with constant weights $w_j = 2\pi/M$ where $j=1,\dots,M$.

Due to the cylindrical symmetry, the kernel functions depend only on the difference of
the angular components $\Delta\varphi=\varphi'-\varphi$.  Because the weights
of the trapezoidal rule are constant and its nodes proportional to the indices,
the discretized block matrix then depends through the difference of the angles
only on the difference of the indices, i.e. $\Delta\varphi_{i,j} =
\Delta\varphi_{i-j}$. Such a block matrix is called circulant and can be
block-diagonalized by a discrete Fourier transform. The blocks on the diagonal
then correspond to the contributions for each angular momentum index starting
from $m=0,\pm1,\dots$ up to $\pm (M-1)/2$ when $M$ is odd or $M/2$ when $M$ is
even.  Note that opposite signs in the angular momentum index contribute
equally. This is due to the fact that angular momentum indices of opposite
signs are connected through the Fourier transform by the transformation
$\Delta\varphi \rightarrow -\Delta\varphi$. Such a transformation, however,
leaves the van der Waals interaction unchanged, since it corresponds to a flip
in the sign of the $z$-coordinates.

The reflection operators of the two objects may in general be non-diagonal as
it is the case in the geometry of two spheres.  In Sec.~\ref{sec:2} it was
argued that the discretized round-trip matrix can then be written in terms of a
product of two block matrices.  In order to exploit the cylindrical symmetry,
the quadrature rule of the angular component of the $\mathbf{k}''$-integral in
\eqref{eq:general-kernel} needs to be a trapezoidal rule of the same order $M$
as above. Only then both block matrices become circulant such that
after the discrete Fourier transform their block matrix product can be
simplified to a product of block-diagonal matrices.

It is possible to perform the discrete Fourier transform analytically, which
opens the possibility for a hybrid numerical method in which the matrix
elements are constructed by discretizing the radial transverse momentum $k$ for
each angular momentum index $m$. At first sight, one might favor such a hybrid
approach over the pure plane-wave approach discussed above, since one can save
on the computation time of the discrete Fourier transform.  In practice,
however, the time needed for the discrete Fourier transform in the
plane-wave approach is dominated by the computation of the matrix elements.
Numerical tests on the plane-sphere geometry indicate that the hybrid approach
is slower than the plane-wave approach.

\section{Application to geometries involving spheres}
\label{sec:3}
The plane-wave method described in the previous section will now be applied to
colloidal systems involving spheres. First, we discuss the scattering of plane
waves at a sphere. Then we study the plane-wave method for the plane-sphere and
the sphere-sphere geometries. Both scattering geometries are cylindrically
symmetric so that the simplification discussed in
Sec.~\ref{sec:cylindrical_symmetry} applies.

\subsection{Plane-wave scattering at a sphere}

For a sphere of radius $R$, the kernel of the reflection operator is given by
\begin{equation}\label{eq:kernel-reflection}
\begin{aligned}
K_{\mathcal{R}_\mathrm{S}}(k,\varphi,\mathrm{TM}; k',\varphi',\mathrm{TM})
	&= \frac{2 \pi \lambdabar_m k}{\kappa}\left[A S_2 + B S_1 \right] \\
K_{\mathcal{R}_\mathrm{S}}(k,\varphi,\mathrm{TE}; k',\varphi',\mathrm{TE})
	&= \frac{2 \pi \lambdabar_m k}{\kappa}\left[A S_1 + B S_2 \right] \\
K_{\mathcal{R}_\mathrm{S}}(k,\varphi,\mathrm{TM}; k',\varphi',\mathrm{TE})
	&= -\frac{2 \pi \lambdabar_m k}{\kappa}\left[C S_1 + D S_2 \right] \\
K_{\mathcal{R}_\mathrm{S}}(k,\varphi,\mathrm{TE}; k',\varphi',\mathrm{TM})
	&= \frac{2 \pi \lambdabar_m k}{\kappa}\left[C S_2 + D S_1 \right]
\end{aligned}
\end{equation}
with the imaginary angular wavelength in the medium
\begin{equation}
 \lambdabar_m = \frac{c}{\sqrt{\epsilon_m}\xi}\,.
\end{equation}
Explicit expressions for the Mie scattering amplitudes $S_1$ and $S_2$ are
given below. As these scattering amplitudes are taken with respect to a
polarization basis referring to the scattering plane, the rotation into the
polarization basis of TE and TM modes defined with respect to the $z$-axis
gives rise to the coefficients $A$, $B$, $C$, and $D$ specified in
Appendix~\ref{sec:appa}. For convenience, the factor $k$ arising from the
integration measure in polar coordinates has been absorbed into the kernel
functions.

The Mie scattering amplitudes for plane waves with polarization perpendicular
and parallel to the scattering plane are defined through an expansion in terms
of partial waves with angular momentum $\ell$ as \citep{Kvien1998,BH}
\begin{equation}
\label{eq:SA}
\begin{aligned}
S_1 &= \sum_{\ell=1}^\infty \frac{2\ell+1}{\ell(\ell+1)}
    \left[ a_\ell\pi_\ell\big(\cos(\Theta)\big) + b_\ell\tau_\ell\big(\cos(\Theta)\big)\right]\\
S_2 &= \sum_{\ell=1}^\infty \frac{2\ell+1}{\ell(\ell+1)}
    \left[ a_\ell\tau_\ell\big(\cos(\Theta)\big) + b_\ell\pi_\ell\big(\cos(\Theta)\big)\right]\,,
\end{aligned}
\end{equation}
respectively. The Mie coefficients $a_\ell$ and $b_\ell$ for electric and magnetic
polarizations, respectively, depend on the electromagnetic response of the homogeneous
sphere and are evaluated at the imaginary size parameter $R/\lambdabar_m$.

The angle $\Theta$ between the directions of the outgoing and incoming wave
is given through
\begin{equation}
\cos(\Theta) = -\lambdabar_m^2 \left[k k' \cos(\varphi-\varphi') + \kappa\kappa'\right] \,.
\label{eq:costheta}
\end{equation}
The angular functions \cite{BH}
\begin{equation}
\begin{aligned}
\pi_\ell(z) &= {P_\ell}^\prime(z) \\
\tau_\ell(z) &= -(1-z^2){P_\ell}^{\prime\prime}(z)+z{P_\ell}^\prime(z)\,,
\end{aligned}
\label{eq:pi_tau}
\end{equation}
are expressed in terms of Legendre polynomials $P_\ell(z)$ and the prime indicates
a derivative with respect to the argument $z=\cos(\Theta)$.

\subsection{Plane-sphere geometry}

For a sphere with radius $R$ above a plane at a surface-to-surface
distance $L$, the kernel of the round-trip operator is given by
\begin{equation}\label{eq:new-kernel}
K_\mathcal{M}(\mathbf{k}, p; \mathbf{k'}, p') =r_p e^{-(\kappa+\kappa')(L+R)} K_{\mathcal{R}_\mathrm{S}}(\mathbf{k}, p; \mathbf{k'}, p')
\end{equation}
with the Fresnel coefficients $r_p$ given in Appendix~\ref{appendix:fresnel} and the kernel of the reflection
operator at the sphere $K_{\mathcal{R}_\mathrm{S}}$ defined in Eq.~\eqref{eq:kernel-reflection}.
The exponential function corresponds to the translation of the plane waves from the plane to the
sphere center and back.

In the multipole method, a symmetrization of the round-trip operator is
important for a fast and stable numerical evaluation of the van der Waals interaction
because otherwise the matrices appearing in the calculation are
ill-conditioned.\cite{Hartmann2018}  This symmetrization is not as crucial in
the plane-wave method where it merely gives a factor of two in run-time speed-up
because only half of the matrix elements need to be computed.

It is however important to write the translation over the sphere radius in
\eqref{eq:new-kernel} symmetrically with respect to the two momenta $\kappa$
and $\kappa'$. Only then the matrix elements in the plane-wave method are
well-conditioned and take their maximum around $\mathbf{k}=\mathbf{k}'$.  This can be
understood by examining the asymptotic behavior of the round-trip operator when
$R\gg L$. By employing the asymptotics of the scattering amplitudes for
large radii,\cite{Nussenzveig69, Nussenzveig92} the leading $R$-dependent
contribution of the kernel of the round-trip operator can be identified as the
exponential factor \cite{Spreng2018, Henning2019}
\begin{equation}\label{eq:exponential_factor}
 \exp\left\{-\frac{R}{\lambdabar_m}\left[\lambdabar_m(\kappa+\kappa')
				-2\sin\left(\frac{\Theta}{2}\right)\right]\right\}\,.
\end{equation}
with the angle $\Theta$ defined through (\ref{eq:costheta}).
Its main contribution comes from $k=k'$ and $\varphi=\varphi'$
where the exponent vanishes. When the translation operator is not expressed
symmetrically with respect to the momenta, the kernel would grow exponentially
with $\kappa$ and decrease exponentially with $\kappa'$ or vice versa,
resulting in an ill-conditioned matrix.

\subsubsection{Quadrature rule for radial wave vector component}
\label{sec:radial_quadrature}
Before the van der Waals interaction can be determined numerically, the quadrature
rule for the integration over the radial component of the transverse wave vectors
in \eqref{eq:integraloperatordiscretized} needs to be specified.
In principle, any quadrature rule for the semi-infinite interval
$[0, \infty)$ can be used. The Fourier-Chebyshev scheme described in
Ref.~\onlinecite{Boyd1987} turned out to be particularly well suited.
Defining
\begin{equation}
t_n = \frac{\pi n}{N+1}\,,
\end{equation}
the quadrature rule is specified by its nodes
\begin{equation}\label{eq:quadrature-points}
k_n = b \cot^2(t_n/2)
\end{equation}
and weights
\begin{equation}
w_n = \frac{8 b \sin(t_n)}{[1-\cos (t_n)]^2}\frac{1}{N+1} \sum_{\substack{ j=1 \\\text{$j$ odd}}}^N \frac{\sin(j t_n)}{j}
\end{equation}
for $n=1,\dots,N$. An optimal choice for the
free parameter $b$ can boost the convergence of the computation.

For dimensional reasons, the transverse wave vector and thus $b$ should scale
like the inverse surface-to-surface distance $1/L$. In fact, the choice $b=1/L$
already yields a fast convergence rate and will be used in the following discussion.

\subsubsection{Estimation of the convergence rate}\label{sec:convergence_rate}

In order to understand how well the plane-wave method performs when $R \gg L$,
one needs to know how the quadrature orders $N$ and $M$ for the integration
over the angular and radial wave vector component, respectively, scale with the
aspect ratio $R/L$ for a maximally allowed relative error. We refer to this
scaling as the convergence rate of the quadrature schemes.

In the multipole method, the number of multipoles needs to be truncated in
order to make the calculation of the van der Waals interaction amenable to linear
algebra routines.  Convergence will then be reached when the highest multipole
index $\ell_\mathrm{max}$ and the highest angular momentum index
$m_\mathrm{max}$ included in the computation take values which scale as
$\ell_\mathrm{max}\sim R/L$ and $m_\mathrm{max} \sim \sqrt{R/L}$,
respectively.\cite{Teo2011, Teo2012, Teo2013} Since the angular quadrature
order $M$ in the plane-wave approach is related to $m_\mathrm{max}$ through a
Fourier transform, one can expect that it exhibits the same convergence rate as
$m_\mathrm{max}$.  The scaling behavior of the radial quadrature order $N$,
however, is a priori not known and will be determined in the following.

Within the plane-wave approach, convergence can only be reached once the nodes
associated to the quadrature rules are able to resolve the structure of the
kernel functions.  The important contributions of the round-trip kernel come
from a region around its maximum at $\kappa=\kappa'\approx 1/L$. This
region corresponds to indices of the Fourier-Chebyshev quadrature rule around
$n=n'\approx N/2$, where for large $N$ the spacing between neighboring
quadrature nodes is given by
\begin{equation}
\delta k \approx \frac{2\pi}{L N}\,.
\end{equation}
Furthermore, when $R\gg L$, the kernel can be approximated by a Gaussian with
width $\sqrt{\kappa/R}\sim 1/\sqrt{LR}$. This can be seen by expanding
the exponent in \eqref{eq:exponential_factor} around $k=k'$. Requiring $\delta k$
to be of the order of that Gaussian width, we find that the quadrature order $N$
scales like $\sqrt{R/L}$. Along the same lines, it can be verified that the angular
quadrature order $M$ obeys the same scaling law.

The quadrature orders for angular and radial integration can thus be expressed
as
\begin{equation}\label{eq:etas}
 N = \left\lceil\eta_N \left(\frac{R}{L}\right)^{1/2}\right\rceil\,,\quad
 M = \left\lceil\eta_M \left(\frac{R}{L}\right)^{1/2}\right\rceil\,
\end{equation}
respectively, where the ceiling function ensures that the orders are integers.
The two coefficients $\eta_N$ and $\eta_M$ control the numerical accuracy with
larger values corresponding to higher accuracy.

\begin{figure}
 \includegraphics[width=\columnwidth]{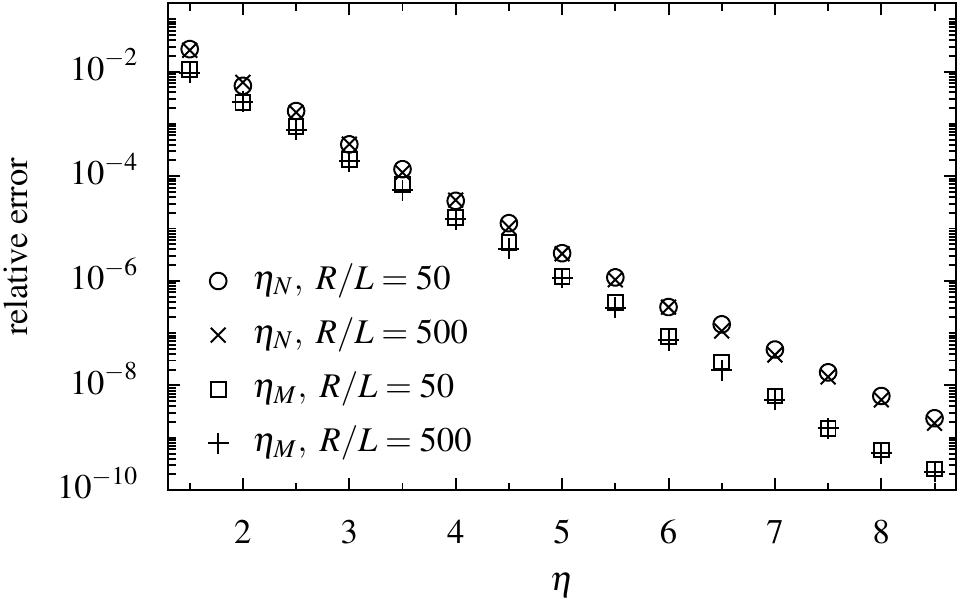}
 \caption{The relative error as a function of $\eta_N$ (circles and crosses)
	and $\eta_M$ (squares and pluses) for the aspect ratios $R/L=50$ and
	$500$.}
 \label{fig:error_vs_eta}
\end{figure}

These expectations for the convergence rate can be verified numerically.  We
specifically consider perfect reflectors in vacuum, a sphere radius of
$1\,\mu{\rm m}$, i.e.\ a typical value for colloids, and room temperature,
$T=293\,{\rm K}$. Figure~\ref{fig:error_vs_eta} shows the relative error of the
van der Waals free energy as a function of $\eta_N$ and $\eta_M$ for the aspect
ratios $R/L=50$ and $500$.  The errors have been computed relative to energies
with much larger values of the coefficients, namely $\eta_N=\eta_M=14$ for all
points in the figure. For the points where the relative errors are shown as a
function of $\eta_N$, the coefficient of the angular quadrature order was kept
fixed at $\eta_M=14$, and vice versa.  One can indeed see that the coefficients
depend only weakly on the aspect ratio $R/L$. This also holds for other system
parameters and materials of the objects. The figure can be further used as a
guide to choose $\eta_N$ and $\eta_M$ in order to obtain a given numerical
accuracy.  The formulas \eqref{eq:etas} only work when $R/L$ is larger than 50.
For smaller aspect ratios, one can simply set $R/L$ to $50$ in
Eq.~\eqref{eq:etas}, which gives a sufficiently high accuracy depending on the
coefficients $\eta_N$ and $\eta_M$.

In comparison to the multipole method we conclude that the matrix sizes
appearing in the plane-wave approach are smaller by a factor of $\sqrt{R/L}$.
This reduction in the matrix size becomes particularly relevant when typical
aspect ratios appearing in experiments are considered.

\subsubsection{Runtime analysis: plane-wave versus multipole method}

We now further quantify the advantages of the plane-wave method over the
multipole method by analyzing their respective runtimes.  The plane-wave method
was implemented in Python using the scientific libraries NumPy,\cite{numpy}
SciPy\cite{scipy} as well as Numba\cite{numba} for just-in-time compilation.
For the multipole method the implementation of Ref.~\onlinecite{HartmannJOSS}
in C was used. Because the latter only supports the computation of the van der
Waals free energy, we restrict the analysis to this quantity.

We consider the same plane-sphere setup as in Sec.~\ref{sec:convergence_rate} with
perfect reflectors in vacuum, a sphere radius of $R=1\,\mu{\rm m}$ and temperature
$T=293\,{\rm K}$.  The van der Waals free energy at finite temperatures can be
evaluated in different ways. We consider the Matsubara spectrum decomposition
(MSD) represented by Eq.~\eqref{eq:F}, and the Pad\'e spectrum decomposition
(PSD) \cite{Hu2010} outlined in Appendix~\ref{appendix:PSD}.  When using PSD,
only $\sqrt{\lambda_T/L}$ terms in the frequency summation need to be
considered, where $\lambda_T=\hbar c/k_BT$ is the thermal wavelength.
Thus, PSD is expected to be significantly faster than MSD
for all experimentally relevant distances since the latter requires
a summation over $\lambda_T/L$ terms to ensure convergence.
\begin{figure}
 \includegraphics[width=\columnwidth]{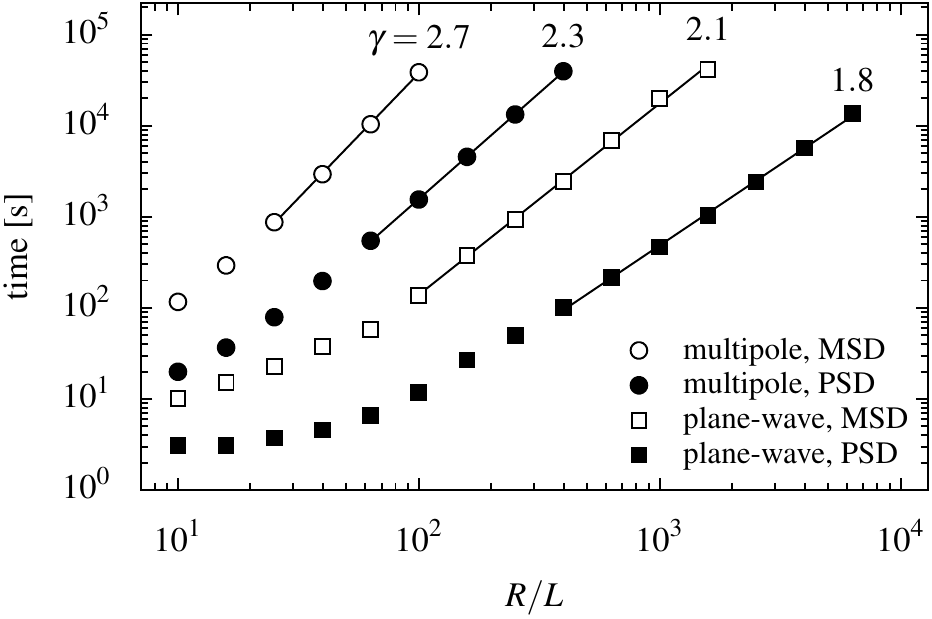}
	\caption{Runtime comparison between the plane-wave method (squares)
	and the multipole method (circles) using MSD (open symbols) and PSD
	(filled symbols). The solid lines indicate a power-law
	fit $\sim (R/L)^\gamma$ on the basis of the data points shown on top
	of the lines. The value of the exponent $\gamma$ is indicated at the
	end of the lines. The timing experiments were carried out
	on a computer with an Intel Core i7-2600 processor. The four cores
	running at $3.4\,{\rm GHz}$ were fully exploited by running eight threads
	or processes in parallel.}
 \label{fig:timing}
\end{figure}

To ensure comparability, the van der Waals free energy is computed with both methods
to the same numerical precision of about six correct digits.
Figure~\ref{fig:timing} shows the runtime of the van der Waals free energy for
the multipole method (circles) and the plane-wave method (squares) for aspect
ratios $R/L\geq10$ using MSD (open symbols) and PSD (filled symbols). For all
timing experiments a machine with an Intel Core i7-2600 processor was used. The
four cores running at $3.4\,{\rm GHz}$ were fully exploited by either running
eight threads or processes in parallel depending on the implementation.
We find that the plane-wave method is significantly faster than the multipole
method.  As expected, the PSD performs better than the MSD.  For instance, at the aspect
ratio $R/L=100$, the multipole methods takes about 11 hours to compute the free
energy using MSD and only 25 minutes with PSD. The plane-wave method, however,
needs only about two minutes to compute the same quantity when using MSD and 12
seconds with PSD.  For other system parameters and materials we
come to similar conclusions for the runtime.

The black lines in Fig.~\ref{fig:timing} are fits to the points they overspan.
The timings of the multipole method are consistent with the timing experiment
in Ref.~\onlinecite{Hartmann2018} where it was found that for a given frequency
and angular momentum index the timing scales as $\sim (R/L)^{1.31}$. The sum
over the angular momentum indices scales with $\sim (R/L)^{0.5}$. The above
mentioned scaling behavior for the frequency sum in the MSD and PSD is thus
consistent with the observed over-all scaling of $(R/L)^{2.7}$ and $(R/L)^{2.3}$,
respectively.

The method based on plane waves scales as $(R/L)^{2.1}$ for MSD and
$(R/L)^{1.8}$ using PSD, allowing the computation of higher aspect ratios with
ordinary hardware.  The difference between the scaling behavior of the MSD and
PSD for the plane-wave method of about $(R/L)^{0.3}$ is notably smaller than
the expected difference of $(R/L)^{0.5}$.  As discussed in appendix
\ref{appendix:PSD}, while the PSD requires the evaluation of fewer frequency
contributions to the Casimir energy, some of the frequencies to be considered
are higher than those required for the MSD. Numerical tests show that the time needed
to evaluate matrix elements increases with increasing frequency, thus offering
an explanation for the reduced improvement of the PSD over the MSD.

Note that for the calculations of the determinants, we did not use the
sophisticated algorithm using hierarchical matrices which was crucial to boost
the performance for Casimir computations in the multipole
basis.\cite{Hartmann2017,Hartmann2018}  Since we are dealing with much smaller
matrices and our computation time is dominated by the calculation of the matrix
elements itself, such method is not expected to bring a significant
improvement. Instead we speed up the calculation of the matrix elements by
first estimating their values in terms of their asymptotic behavior given in
Eq.~\eqref{eq:exponential_factor}.  Since the matrices are well-conditioned and
their dominant contributions come from matrix elements around the diagonal, we
can set matrix-elements to zero if their asymptotic behavior predicts a value
smaller than the machine precision.  Otherwise, the computed matrix elements are
numerically exact. Numerical tests reveal that this scheme yields a
speed-up scaling as $(R/L)^{0.5}$.

\subsection{Sphere-sphere geometry}
Another example of a van der Waals setup with cylindrical symmetry consists of
two spheres with radii $R_1$ and $R_2$. As in the plane-sphere geometry, we
denote the surface-to-surface distance as $L$.  The kernel of the round-trip
operator is of the form \eqref{eq:general-kernel} with the kernel functions of
the reflection operators of the respective spheres given in
\eqref{eq:kernel-reflection}. Note that the sign of the coefficients $C$ and
$D$ differs for the two spheres.

We recall that, because the reflection operator at both objects is now
non-diagonal, a discretization of two integrals over the transverse momenta is
required. Firstly, the discretization of the integral over $\mathbf{k}'$ in
Eq.~\eqref{eq:integraloperator} results in a finite matrix representation of
the round-trip operator. Secondly, the discretization of the integral over
$\mathbf{k}''$ in Eq.~\eqref{eq:general-kernel} allows to express the
round-trip matrix in terms of a product of two block-matrices.  For the radial
components we employ the Fourier-Chebyshev quadrature scheme presented in
Sec.~\ref{sec:radial_quadrature}. The quadrature orders, however, do not need
to coincide and thus we use the quadratures of order $N'$ and $N''$ for the
integrations over $k'$ and $k''$, respectively.  Likewise we employ trapezoidal
rules of order $M'$ and $M''$ for the discretization of the angular components.
In order to exploit the cylindrical symmetry of the problem by means of the
discrete Fourier transform, the quadrature orders $M'$ and $M''$ will be required
to be equal. However, for the sake of the following analysis we assume them to
be different.

The convergence rate of the quadrature orders can be determined with the same
line of reasoning as in section~\ref{sec:convergence_rate}. When the sphere
radii become large, the kernel functions of the reflection operators can be
approximated by Gaussians for which the width is controlled by the respective
radii. The kernel of the round-trip operator is then a convolution of these two
Gaussians, resulting in a Gaussian where the width is controlled by the
effective radius
\begin{equation}
\label{eq:effective-radius}
R_\mathrm{eff} = \frac{R_1R_2}{R_1+R_2}
\end{equation}
instead. We then find the scaling
\begin{equation}
N' \sim M' \sim \sqrt{R_\mathrm{eff}/L}\,.
\end{equation}
The quadrature orders $N''$ and $M''$ can be estimated from the
convolution integral. The integrand is a Gaussian where the two
radii appear as a sum, $R_1+R_2$ and thus the quadrature orders scale as
\begin{equation}
N'' \sim M'' \sim \sqrt{(R_1+R_2)/L}\,.
\end{equation}
Note that in the plane-sphere limit, where one radius is much larger than
the other, the quadrature orders $N'$ and $M'$ become the same as in the
plane-sphere geometry \eqref{eq:etas} where $R=R_1$. The quadrature orders $N''$
and $M''$ then become very large, reflecting the fact that the kernel
functions of the sphere with the larger radius $R_2$ become strongly peaked
around $\mathbf{k}=\mathbf{k}'$ as expected from the reflection properties of a
plane.  A more detailed discussion of this limiting procedure in connection
with the PFA can be found in Ref. \onlinecite{Spreng2018}.

Finally, we need to go back to equal orders $M'$ and $M''$. Since $R_1+R_2 >
R_\mathrm{eff}$, the quadrature order of the trapezoidal rule thus scales as
$M'=M''\sim \sqrt{(R_1+R_2)/L}$ to ensure convergence.

\section{Numerical results for colloids}
\label{sec:4}
The plane-wave method developed in this paper will now be applied to various
colloidal systems suspended in an aqueous electrolyte solution.  In particular,
we will study the interaction between two spherical colloidal particles and the
interaction of such a spherical particle with a plane wall.

For the analysis of colloid experiments the Lifshitz theory is most commonly
employed, where the finite curvature of the spheres is accounted for by the
proximity force approximation (PFA).  Within the PFA the van der Waals free energy is
given by
\begin{equation}\label{eq:E_PFA}
\mathcal{F}_\mathrm{PFA} = 2\pi R_\mathrm{eff} \int_L^\infty dl \mathcal{F}_\mathrm{PP}(l)
\end{equation}
with the effective radius $R_\mathrm{eff}$ defined in \eqref{eq:effective-radius}
and the van der Waals free energy per unit area between two parallel planes
\begin{multline}\label{eq:E_PP}
\mathcal{F}_\mathrm{PP}(L) = \frac{k_BT}{2} \sum_{n=-\infty}^{+\infty} \sum_{p\in\{\mathrm{TE}, \mathrm{TM}\}} \int_{\sqrt{\epsilon_m}\vert \xi_n\vert/c}^\infty \frac{d\kappa}{2\pi}\kappa\\
\times \log\left(1- r_p^{(1)}r_p^{(2)} e^{-2\kappa L}\right)
\end{multline}
where $\epsilon_m$ is the relative permittivity of the medium between the
planes.  The Fresnel coefficients $r_p^{(j)}$ of plane $j$ depend on the
materials and are functions of $\xi_n$ and $\kappa$. Explicit expressions are
given in appendix \ref{appendix:fresnel}. By taking the negative derivative
of \eqref{eq:E_PFA} with respect to $L$, a corresponding expression for the
force can be found
\begin{equation}\label{eq:F_PFA}
F_\mathrm{PFA} = 2\pi R_\mathrm{eff} \mathcal{F}_\mathrm{PP}(L)\,.
\end{equation}
The PFA is an asymptotic result valid only in the limit
$L/R_\mathrm{eff}\rightarrow 0$.\cite{Spreng2018} At finite distances between the surfaces,
there will always be some discrepancy between the exact result and the
PFA.

The material dependence of the van der Waals interaction is often expressed in
terms of the Hamaker constant $A$. \cite{Hamaker1937}  Within Hamaker's
microscopic theory, the non-retarded free energy per unit area for two parallel
planes is given by
\begin{equation}
\mathcal{F}_\mathrm{PP} = - \frac{A}{12 \pi L^2}\,,
\end{equation}
which is only valid for very small distances.  For larger separations of the
planes, retardation can no longer be neglected and the free energy needs to be
computed using Eq.~\eqref{eq:E_PP}. This motivates the definition of an effective
Hamaker parameter\cite{Russel1989}
\begin{equation}\label{eq:eff_Hamaker_plpl}
A_\mathrm{eff}(L) = -12\pi L^2 \mathcal{F}_\mathrm{PP}(L)\,,
\end{equation}
which now has a non-trivial distance dependence through the exact plane-plane
free energy per unit area.
Usually $A_\mathrm{eff}$ is
experimentally
determined by measuring the force $F$ between
spherical surfaces:\cite{Wodka2014}
\begin{equation}\label{eq:eff_Hamaker_spsp}
A_\mathrm{eff}(L) = - \frac{6 L^2}{R_\mathrm{eff}} F\,.
\end{equation}
Since the PFA expression  \eqref{eq:F_PFA} becomes exact in the small distance limit, the two definitions
(\ref{eq:eff_Hamaker_plpl}) and (\ref{eq:eff_Hamaker_spsp}) are equivalent as
far as the Hamaker constant $A = A_\mathrm{eff}(0)$ is concerned. However,
deviations from the PFA result make them differ at finite distances. In the
following, we take the experimentally motivated Eq.~(\ref{eq:eff_Hamaker_spsp})
as our definition of the effective Hamaker parameter.  In addition to the
distance dependence associated to electrodynamical retardation, it also
contains a geometry dependence which often translates into further reduction as
the distance increases.

In the following, we will study colloidal systems involving polystyrene and
mercury. The validity of the PFA will be analyzed for the plane-sphere and
sphere-sphere geometry using the exactly calculated van der Waals interaction through
the numerical method developed above. Moreover, the geometry dependence
of the effective Hamaker parameter (\ref{eq:eff_Hamaker_spsp}) will be analyzed.

In our analysis, we will consider the two extreme cases of very low and very
high salt concentrations in the aqueous suspensions.  Only the zero-frequency
Matsubara contribution is affected by the presence of ions in solution, since
the corresponding plasma frequency is many orders of magnitude smaller than
$k_BT/\hbar$ even when considering the highest possible concentrations.  We
follow the standard theoretical modeling of van der Waals screening and
consider the zero-frequency contribution to be completely suppressed by ionic
screening in the case of high salt
concentrations.\cite{Mitchell1974,MahantyNinham1976,Parsegian2005} On the other
hand, we model very low salt concentrations by summing over all Matsubara
frequencies $\xi_n$ including $n=0$ and neglecting the effect of ions on the
dielectric permittivities.  Based on the scattering theory, an alternative
result for the van der Waals interaction in electrolytes has been
derived.\cite{MaiaNeto2019}  This approach will not be discussed here further.

Unless stated otherwise, we model the dielectric function of the materials in
terms of Lorentz oscillators with parameters taken from
Ref.~\onlinecite{Zwol2010}. For polystyrene, we use data set 1 from that
reference. The dielectric function of water was modified to match the correct
static value of $78.7$.  Moreover, the temperature is assumed to be
$T=293\,\mathrm{K}$.

\subsection{Polystyrene in water}
\label{sec:ps_water}

The van der Waals interaction between a polystyrene bead and a glass wall in an
aqueous solution has been experimentally studied using the method of total
internal reflection microscopy.\cite{Bevan1999,Hansen2005} With the colloidal
probe technique, the interaction force between two latex spheres was measured.
\cite{Wodka2014} Based on calculations presented in Refs.~\onlinecite{Pailthorpe1982,
Russel1989}, Elzbieciak-Wodka \textit{et al.}\ assumed that the accuracy of the
PFA for particles with diameters above $0.5\,\mu \mathrm{m}$ up to separations
of $100\,\mathrm{nm}$ is within 1\%. Deviations of the measured forces from the PFA result,
which resulted in a smaller Hamaker constant, were
attributed to the surface roughness of the spheres.  Motivated by these
experiments, we use the numerical method developed in this paper to study the
van der Waals interaction between two polystyrene bodies in an aqueous solution for
the plane-sphere and sphere-sphere geometry.

\begin{figure*}
 \includegraphics[width=\textwidth]{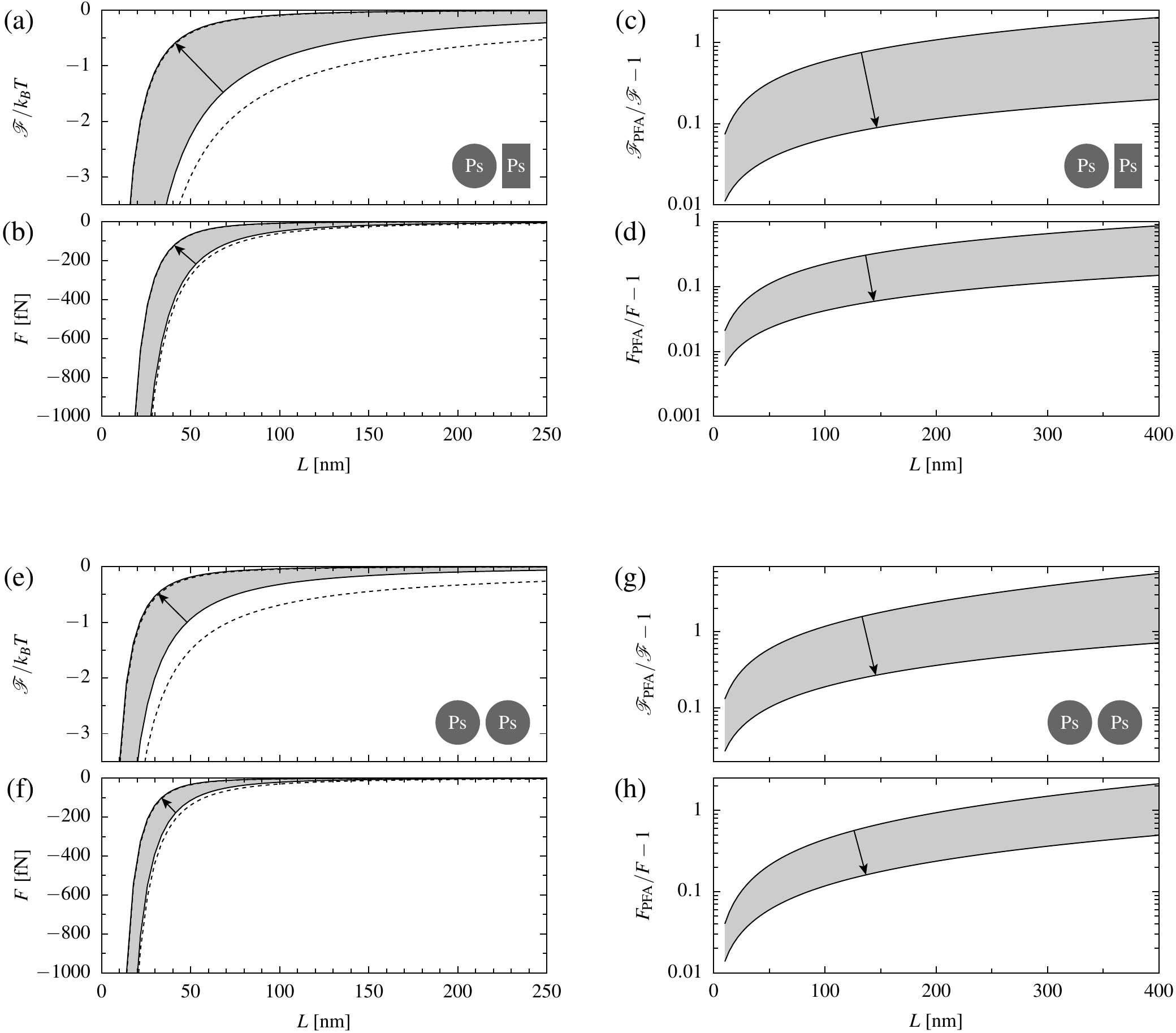}
 \caption{(a) van der Waals free energy and (b) force for a polystyrene
 sphere with radius $R=1\,\mu\mathrm{m}$ in front of a polystyrene plane in water
 as a function of the surface-to-surface distance $L$. Solid lines correspond to
 the numerically exact values and dashed lines to the PFA. The arrow indicates
 the direction of increasing screening so that the lower curve includes the
 Matsubara frequency $\xi_0$ while the upper curve does not. The grey shaded
 area indicates the interaction for any intermediate screening strength. The
 corresponding relative error of the PFA for (c) the van der Waals free energy and
 (d) force. Here, the upper curve corresponds to the absence of screening
 and screening increases through the grey area as indicated by the arrow.
 (e) van der Waals free energy and (f) force for two polystyrene spheres with radii
 $R_1=R_2=1\,\mu\mathrm{m}$ in water as a function of the surface-to-surface
 distance $L$. Corresponding relative error of the PFA for (g) the free energy and (h)
 the force.}
 \label{fig:ps_ps_water}
\end{figure*}

The van der Waals free energy and force between a plane and a sphere with
radius $R=1\,\mu\mathrm{m}$ as a function of the surface-to-surface distance
$L$ is depicted in Figs.~\ref{fig:ps_ps_water} (a) and (b), respectively. The
solid lines represent the numerically exact values, while the dashed lines
correspond to the PFA.  Here and in the following figures, the arrow indicates
the direction of increasing screening. Thus, here, the upper curves represent
strong screening while the lower curves refer to no screening.  The van der Waals
interaction for intermediate screening then will follow a curve in the grey shaded
area. For both observables, the PFA overestimates the interaction and the
approximation agrees better with the exact values when the screening is
strong.

The relative error of the PFA for the van der Waals free energy and the van der
Waals force is quantified in Figs.~\ref{fig:ps_ps_water} (c) and (d),
respectively.  We find that the PFA is more accurate for the force than for the
free energy.  The relative error of the PFA is larger than one percent above a
separation of about $10\,\mathrm{nm}$ for the energy and above about
$20\,\mathrm{nm}$ for the force regardless of the screening strength.  The PFA
performs worse when screening is negligible because the corrections to the PFA are
particularly large for the zero-frequency contribution.  In fact, in view of the
derivative-expansion approach, it is expected that the short-distance expansion
of the zero-frequency contribution to the free energy contains corrections to
the PFA which are logarithmic in $L/R$ and thus
particularly large at small separations.\cite{BimonteAPL2012}

The van der Waals free energy and force for two polystyrene spheres
with equal radii $R_1=R_2=1\,\mu\mathrm{m}$ as a function of the surface-to-surface distance $L$
is depicted in Figs.~\ref{fig:ps_ps_water} (e) and (f), respectively.
Again, the PFA overestimates
the van der Waals interaction and performs better when screening is strong.
Overall, the free energy and the force are smaller for the two spheres than for
the plane and the sphere. This can be explained by the fact that the effective
interacting surface area is smaller in the former than in the
latter.\cite{Spreng2018}

The relative error of the PFA for the van der Waals free energy and for the van der Waals
force are shown in Figs.~\ref{fig:ps_ps_water} (g) and (h), respectively.  Similar
as in the plane-sphere geometry, the accuracy of the PFA is better when
screening is strong and the PFA is more accurate for the force than for
the energy.  Above separations of $10\,\mathrm{nm}$, the relative
error is larger than 1\% for any screening strength. This is in
particular true for distances below $100\,\mathrm{nm}$, which is in contradiction
to the assumption made in Ref.~\onlinecite{Wodka2014}. Compared to the
plane-sphere geometry, the PFA is less accurate for two spheres.
This is consistent with the fact that in the plane-sphere geometry the correction to PFA is dominated
by diffractive contributions. \cite{Henning2019} When considering two spheres,
these diffractive contributions add up and lead to a larger correction to the PFA.

\begin{figure}
 \includegraphics[width=\columnwidth]{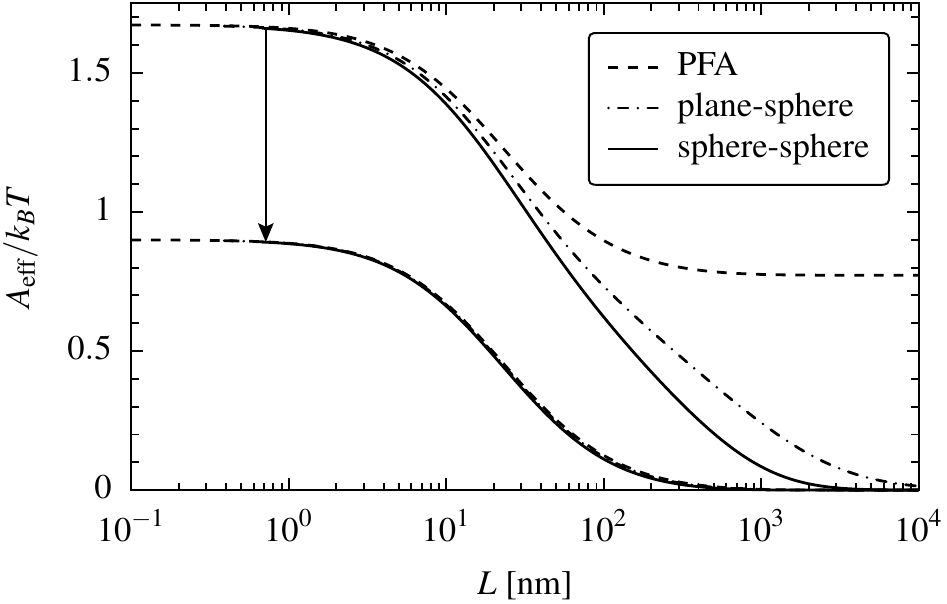}
 \caption{Effective Hamaker parameter for two polystyrene objects in water
 as a function of the surface-to-surface distance $L$. The dash-dotted and
 solid lines are derived from the exact plane-sphere and sphere-sphere
 interaction forces, respectively.  The dashed line is computed within the PFA and is the same for both geometries.
 The radius of the sphere(s) is $1\,\mathrm{\mu m}$.
 The arrow indicates the direction
 of increasing screening strength with the upper (lower) curve including
 (excluding) the Matsubara frequency $\xi_0$.}
 \label{fig:hamaker}
\end{figure}

The effective Hamaker parameter
for spherical surfaces is determined by Eq.~\eqref{eq:eff_Hamaker_spsp}
and depends not only on the chosen
materials but also on the geometry used in its derivation.
Figure~\ref{fig:hamaker} demonstrates this dependence for polystyrene and
water.  The dash-dotted and the solid  curves represent the exact
effective Hamaker parameter for the plane-sphere and sphere-sphere geometries,
respectively, whereas the dashed curve corresponds to the PFA result, which
is  the same for both geometries.
The upper curves do not take screening into account as they
include the full contribution from the Matsubara frequency $\xi_0.$
In contrast, in the lower curves the Matsubara frequency $\xi_0$ is omitted so that these curves correspond to the limit of
a vanishingly small Debye screening length: $\lambda_D\rightarrow 0.$
For any finite value of $\lambda_D,$ the Hamaker parameter for each geometry
first starts close to the upper curve at short distances ($L\ll \lambda_D$) and then is further suppressed by screening,
approaching the lower curve for $L\gg \lambda_D.$

At small separations, the effective Hamaker parameters derived for the two different
geometries asymptotically approach each other and the PFA curve as expected.
As the distance increases,
the reduction of the effective Hamaker parameter calculated within the PFA accounts for electrodynamical retardation only,
whereas the exact curves display an additional reduction associated to curvature.
Such geometrical reduction is more
apparent in the absence of screening, since in this case the PFA curve at long distances
defines a plateau associated to the contribution from the Matsubara frequency $\xi_0,$ while
the exact values decay to zero due to the curvature suppression.

We obtain the value  $A=A_\mathrm{eff}(L\rightarrow 0)=1.67\,k_BT$ for the Hamaker constant.
The limiting value is obtained from
the short-distance plateau defined by the upper curves in Fig.~\ref{fig:hamaker} since they correspond to $L\ll \lambda_D.$
On the other hand, the short-distance plateau associated to the lower curves yields the value $0.90\,k_BT,$
representing the difference between  the Hamaker constant and the contribution from the Matsubara frequency $\xi_0,$
which in turn corresponds to the long-distance plateau defined by the upper PFA curve in Fig.~\ref{fig:hamaker}.
Our value for $A$ is about half of the theoretical value found in the literature.\cite{Russel1989}
This is because the optical data in Ref.~\onlinecite{Zwol2010} used here differ
from Parsegian's optical data set.\cite{Parsegian2005} It is interesting to
observe that even though the difference between the permittivities of the two
data sets is small, namely less than 10\%, the difference for the resulting
Hamaker constants can be much bigger. This is because, at least within the PFA,
the permittivities of the objects and the medium enter in terms of their
differences.  For polystyrene and water, the optical data almost match for UV
frequencies. These frequencies become more and more important as the distance
between the surfaces decreases, and then small differences in the optical data
can result in relatively large differences in the Hamaker constant. The
reduction of the Hamaker constant observed in the experiment of
Ref.~\onlinecite{Wodka2014} could hence be partly due to uncertainties of the
optical data.

\subsection{Mercury and polystyrene in water}

\begin{figure}
 \includegraphics[width=\columnwidth]{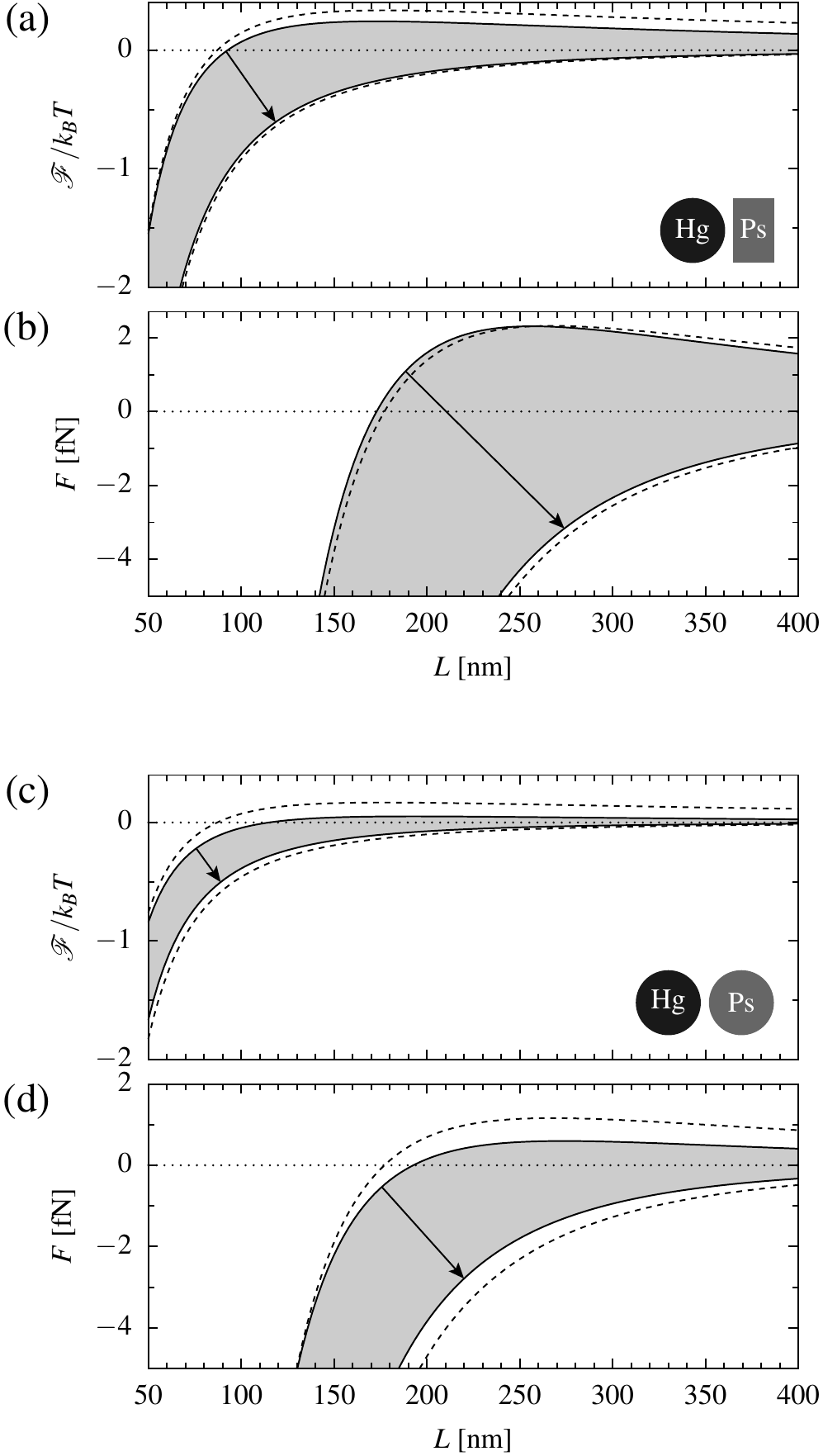}
 \caption{(a) van der Waals free energy and (b) force between a mercury sphere of
 radius $R=1\,\mu\mathrm{m}$ and a polystyrene plane in water.  (c) van der Waals
 free energy and (d) force between a mercury sphere and a polystyrene sphere
 with radii $R_1=R_2=1\,\mu\mathrm{m}$ in water. For both scattering
 geometries, the quantities are presented as a function of the
 surface-to-surface distance $L$.  The solid lines correspond to the numerically
 exact results and the dashed lines to the PFA.  A positive (negative) sign of
 the force represents repulsion (attraction). The arrows indicate the direction
 of increasing screening strength with the upper (lower) curve including
 (excluding) the Matsubara frequency $\xi_0$. The grey shaded area indicates
 the interaction for any intermediate screening.}
 \label{fig:hg_ps_water}
\end{figure}

Mercury and polystyrene in an aqueous medium constitute an interesting colloid
system, since the van der Waals force can be tuned from repulsion to attraction
depending on the screening of the zero frequency contribution.\cite{Ether2015}
Furthermore, due to the high surface
tension mercury droplets have a small surface roughness and, thus, corrections due
to roughness may play a minor role.\cite{Esquivel-Sirvent2014}

We study the interaction between a mercury droplet with radius $R=1\,\mathrm{\mu
m}$ and a polystyrene wall and the interaction between a mercury droplet with a
polystyrene sphere with equal radii $R_1=R_2=1\,\mathrm{\mu m}$.  The dielectric
function of mercury is described by the Drude-Smith model \cite{Smith2001} with
parameters taken from Ref.~\onlinecite{Esquivel-Sirvent2014}.
Figures~\ref{fig:hg_ps_water} (a) and (b) depict the van der Waals free energy and
force in the plane-sphere geometry, respectively. The corresponding
quantities in the sphere-sphere geometry are shown in
Figs.~\ref{fig:hg_ps_water} (c) and (d).\cite{remark} Solid lines correspond to
the numerically exact results, and the dashed lines to the PFA. We use the
convention that a negative sign of the force corresponds to attraction and a
positive sign corresponds to repulsion.

When screening is strong, the free energy and the force are negative and
monotonic. For negligible screening, both quantities are non-monotonic and can change
their sign.  At intermediate distances, the force can be tuned from attractive
to repulsive depending on the screening strength.  Consistent with the
discussion of polystyrene in water, the PFA is more accurate in the
plane-sphere geometry than in the geometry of two spheres. This becomes most
evident when considering the points in which the observables vanish. For
instance, according to the PFA the force vanishes at about $L=178\,\mathrm{nm}$
for both geometries. The equilibrium distance is overestimated by about
$4\,\mathrm{nm}$ for the plane-sphere geometry and underestimated by about
$16\,\mathrm{nm}$ for the two spheres.
The determination of the equilibrium distance is particularly relevant
for stable equilibria. This is the case for the materials considered in connection with
 ice particles \cite{Thiyam2018} and gas bubbles \cite{Esteso2019} in liquid water near a planar interface.
 Our results suggest that beyond-PFA corrections in the nm range could appear when considering
 aspect ratios comparable to those taken in Fig.~\ref{fig:hg_ps_water}.

\begin{figure}
 \includegraphics[width=\columnwidth]{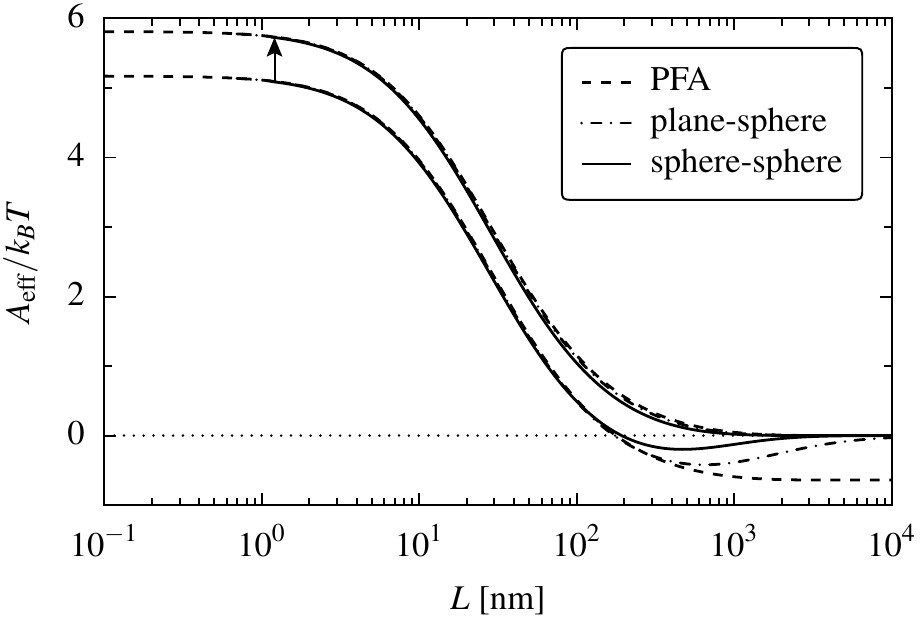}
 \caption{Effective Hamaker parameter for mercury and polystyrene in water
 as a function of the surface-to-surface distance $L$. The dash-dotted and
 solid lines are derived from the exact plane-sphere and sphere-sphere
 interaction forces, respectively.  The dashed line is computed within the PFA and is the same for both geometries.
 The radius of the sphere(s) is $1\,\mathrm{\mu m}$.
 The arrow indicates the direction
 of increasing screening strength with the lower (upper) curve including
 (excluding) the Matsubara frequency $\xi_0$.}
 \label{fig:hamaker_hg_ps_water}
\end{figure}

Figure~\ref{fig:hamaker_hg_ps_water} shows the effective Hamaker parameter for
mercury and polystyrene in water. The effective Hamaker parameter has been
computed through the exact
force between a sphere and a plane (dash-dotted lines) and the exact force
between two spheres (solid lines).
We also show the results obtained within the PFA (dashed lines), which are the same for both geometries.
The contribution from the Matsubara frequency $\xi_0$ is included in the lower lines but not
in the upper ones. For any given Debye screening length,
the Hamaker parameter exhibits a crossover from
the lower curve to the upper one as the distance increases past $\lambda_D.$
We find $A=A_\mathrm{eff}(L\rightarrow 0)=5.17\,k_BT$ for the Hamaker constant by following the lower short-distance plateau.
 In this configuration, the contribution $5.81\,k_BT$ from non-zero Matsubara frequencies,
 associated to the short-distance upper plateau,
is larger than the Hamaker constant.
This is a consequence of the repulsive nature of the contribution from the Matsubara frequency $\xi_0,$
which corresponds to the negative plateau defined by the lower PFA curve for the longer distances shown in Fig.~\ref{fig:hamaker_hg_ps_water}.

Again, the modification of the effective Hamaker parameter associated to the scattering
geometry is most pronounced for larger distances provided that screening is negligible.
The corresponding exact curves exhibit  a non-monotonic behavior as they
tend to zero at large distances.

\section{Conclusions}
\label{sec:5}
A numerical scheme for computing the van der Waals interaction based on the
plane-wave representation of the fluctuating electromagnetic modes was
proposed.  After a Nystr\"om discretization of the plane-wave momenta, the
scattering operator becomes a finite matrix and standard linear algebra
procedures can be employed. The method is applicable to arbitrary scattering
geometries for which the reflection operators of the individual objects are
known in the plane-wave basis. It was demonstrated that a rotational symmetry
can be exploited by means of a discrete Fourier transform.

The plane-wave numerical method was applied to the plane-sphere and the
sphere-sphere geometry. We found that the method converges faster than the
common method based on spherical multipoles. By conducting a runtime analysis
for the plane-sphere geometry we demonstrated that the new plane-wave method
outperforms a state-of-the-art implementation built on spherical multipoles.

For the van der Waals interaction at finite temperatures, the runtime analysis
included a comparison of the conventional Matsubara summation with an
alternative summation scheme based on a Pad\'e spectrum decomposition. The
latter shows an improved convergence rate resulting in a significant
speed-up for the computation of the van der Waals interaction at small
distances. The plane-wave approach together with the Pad\'e spectrum decomposition
put the numerically exact computation of the van der Waals interaction at
experimentally relevant distances within reach of standard desktop computers.

As an application, we employed the new method to study the accuracy of the PFA
in aqueous colloidal systems. The two extreme cases of very strong and very
weak screening were modeled by excluding and including the zero-frequency
contribution of the van der Waals interaction, respectively. For polystyrene in
water we found that the relative error incurred with the PFA as compared to the
numerically exact evaluation is larger than usually anticipated in the
literature, especially for low salt concentrations. Moreover, this effect is more
pronounced for the interaction of two spheres than for a plane and a sphere.
One important consequence is a geometry-dependent reduction of
the effective Hamaker parameter as the distance increases, which adds to the reduction
effects associated to electrodynamical retardation and screening by ions in solution.

In addition, we studied the van der Waals interaction of a mercury sphere with
a polystyrene sphere or a polystyrene wall. These systems have the interesting
feature that the interaction force can be tuned from repulsive to attractive as
a function of the salt concentration. While for strong screening the force is
always attractive, it is repulsive in the case of negligible screening provided the distance
between the objects is not too small.  In this case,
 the exact effective Hamaker parameter exhibits a non-monotonic distance dependence that results from
a beyond-PFA competition between the repulsive and attractive contributions to the interaction force.

\begin{acknowledgments}
The authors are grateful to Michael Hartmann for many inspiring discussions.
B.S. would like to thank Riccardo Messina for a stimulating
conversation. This work has been supported by the Coordination for the
Improvement of Higher Education Personnel (CAPES) and the German
Academic Exchange Service (DAAD) through the PROBRAL collaboration
program. B.S. was also financially supported by the German Academic
Exchange Service (DAAD) through an individual fellowship. P.A.M.N.
thanks the Brazilian agencies National Council for Scientific and
Technological Development (CNPq), the National Institute of Science and
Technology Complex Fluids (INCT-FCx), the Carlos Chagas Filho
Foundation for Research Support of Rio de Janeiro (FAPERJ), and the São
Paulo Research Foundation (FAPESP).
\end{acknowledgments}

\section*{AIP Publishing Data Sharing Policy}

The data represented in Figs.~\ref{fig:error_vs_eta} to
\ref{fig:hamaker_hg_ps_water} are openly available on Zenodo at
\url{http://doi.org/10.5281/zenodo.3751295}, cf.~Ref.~\onlinecite{DataAtZenodo}.

\appendix
\section{Polarization transformation coefficients}
\label{sec:appa}

The coefficients appearing in \eqref{eq:kernel-reflection} arise
from the transformation between the polarization basis defined with
respect to the scattering plane and the TE/TM polarization basis defined
with respect to the symmetry axis of the setup. They have been derived
in Ref.~\onlinecite{Spreng2018} and are given by
\begin{equation}
\begin{aligned}
A &=
\frac{\cos(\Delta\varphi)-\lambdabar_m^4\big[\kappa\kappa'+kk'\cos(\Delta\varphi)\big]
\big[kk'+\kappa\kappa'\cos(\Delta\varphi)\big]}
{1-\lambdabar_m^4\big[\kappa\kappa'+kk'\cos(\Delta\varphi)\big]^2} \\
B & = -\frac{\lambdabar_m^2 kk'\sin^2(\Delta\varphi)}
{1-\lambdabar_m^4\big[\kappa\kappa'+kk'\cos(\Delta\varphi)\big]^2} \\
C  & = \pm\frac{\lambdabar_m^3[\kappa'k^2+\kappa kk'\cos(\Delta\varphi)]}
{1-\lambdabar_m^4\big[\kappa\kappa'+kk'\cos(\Delta\varphi)\big]^2}\sin(\Delta\varphi) \\
D & = \mp\frac{\lambdabar_m^3[\kappa{k'}^2+\kappa'kk'\cos(\varphi)]}
{1-\lambdabar_m^4\big[\kappa\kappa'+kk'\cos(\Delta\varphi)\big]^2}\sin(\Delta\varphi)\,.
\end{aligned}
\end{equation}
Here, $\Delta\varphi = \varphi'-\varphi$ and the upper (lower) sign in the
coefficients $C$ and $D$ corresponds to an incoming plane wave traveling in
positive (negative) $z$-direction.

\section{Fresnel coefficients}
\label{appendix:fresnel}

The reflection at the interface between two homogeneous half spaces filled with
a medium and a dielectric material with permittivities $\epsilon_m$
and $\epsilon_d$, respectively, is described by the Fresnel coefficients
\cite{BH}
\begin{align}
r_\mathrm{TE}(i\xi, \kappa) &= \frac{c\kappa-\sqrt{c^2 \kappa^2 + \xi^2[\epsilon(i\xi)-1]}}{c\kappa+\sqrt{c^2 \kappa^2 + \xi^2[\epsilon(i\xi)-1]}}\,,\\
r_\mathrm{TM}(i\xi, \kappa) &= \frac{\epsilon(i\xi)c\kappa-\sqrt{c^2 \kappa^2 + \xi^2[\epsilon(i\xi)-1]}}{\epsilon(i\xi)c\kappa+\sqrt{c^2 \kappa^2 + \xi^2[\epsilon(i\xi)-1]}}
\end{align}
with $\epsilon=\epsilon_d/\epsilon_m$.

\section{Pad\'e spectrum decomposition}
\label{appendix:PSD}
Integrals containing Bose or Fermi distribution functions can often be
transformed into sum-over-poles expressions using Cauchy's residue theorem.
The Pad\'e spectrum decomposition (PSD) is a particularly efficient
sum-over-poles scheme,\cite{Hu2010} which can also be used for the computation of
the van der Waals interaction at finite temperatures.  Before explaining the PSD, we
outline the more commonly used scheme involving the Matsubara spectrum
decomposition (MSD). For simplicity, we restrict the discussion to the van der Waals
free energy.

Before using any of the mentioned spectrum decomposition schemes, the van der Waals
free energy at temperature $T$ can be expressed in terms of an integral over
real frequencies \cite{Jaekel1991,Genet2000}
\begin{equation}\label{eq:energy_realfrequencies}
\mathcal{F} = \frac{\hbar}{2\pi} \int_0^\infty d\omega\, f\left(\frac{\hbar\omega}{k_BT}\right)
	\mathrm{Im}\left[\Phi(\omega) \right]
\end{equation}
with
\begin{equation}
\Phi(\omega)=\log\det(1-\mathcal{M}(\omega))
\end{equation}
where the round-trip operator $\mathcal{M}$ is defined in Eq.
\eqref{eq:roundtrip_operator}. Note that here $\mathcal{M}$ is a function of
real frequencies, while in Eq.~\eqref{eq:F} its argument are imaginary
frequencies. For simplicity, we keep the same notation for both operators.  The
temperature dependence of the van der Waals free energy is captured by the function
\begin{equation}
f(x) = 1 + 2 \bar{n}(x)\,,\quad \bar{n}(x) = \frac{1}{e^x-1}\,.
\end{equation}
Besides quantum fluctuations, the function $f$ accounts for thermal
fluctuations through the mean number $\bar{n}$ of photons per mode.  It has
equally spaced poles along the imaginary axis at the Matsubara frequencies. Using
Cauchy's residue theorem, the van der Waals free energy
\eqref{eq:energy_realfrequencies} can be cast into a sum over the imaginary
Matsubara frequencies as given in Eq.~\eqref{eq:F}. This procedure has
subtleties in connection with the zero-frequency
contribution,\cite{Guerout2014} which shall not be discussed here further.  In
numerical applications, the Matsubara sum needs to be truncated and convergence
is reached when the number of terms is of the order of $\lambda_T/L$ with the
thermal wavelength $\lambda_T = \hbar c/k_BT$.

For the PSD, one starts out by expanding the function $f$ in terms of a Pad\'e
approximation,
\begin{equation}
f(x) \approx \frac{2}{x} + 2 x \frac{P_{N-1}(x^2)}{Q_N(x^2)}
\end{equation}
where $P_{N-1}$ and $Q_N$ are polynomials of order $N-1$ and $N$, respectively.
Alternatively, $f$ can be expressed in terms of a sum over its simple poles
\begin{equation}\label{eq:PSD}
f(x) \approx \frac{2}{x} + 1 + 2 \sum_{j=1}^N \left(\frac{\eta_j}{x+i\xi_j} + \frac{\eta_j}{x-i\xi_j}\right)\,,
\end{equation}
where the PSD frequencies $\xi_j$ are determined by the roots of $Q_N$ and can
be computed in terms of eigenvalues of a symmetric tridiagonal matrix. The PSD
coefficients $\eta_j$ can be calculated recursively as detailed in
Ref.~\onlinecite{Hu2010}. The pole in \eqref{eq:PSD} at $x=0$ remains unchanged
with respect to the MSD. Thus, when using the residue theorem, subtleties in
connection with the zero-frequency contribution can be treated in the same
manner as in the MSD.  All other poles now lie unevenly distributed on the
imaginary axis.

\begin{figure}
 \includegraphics[width=\columnwidth]{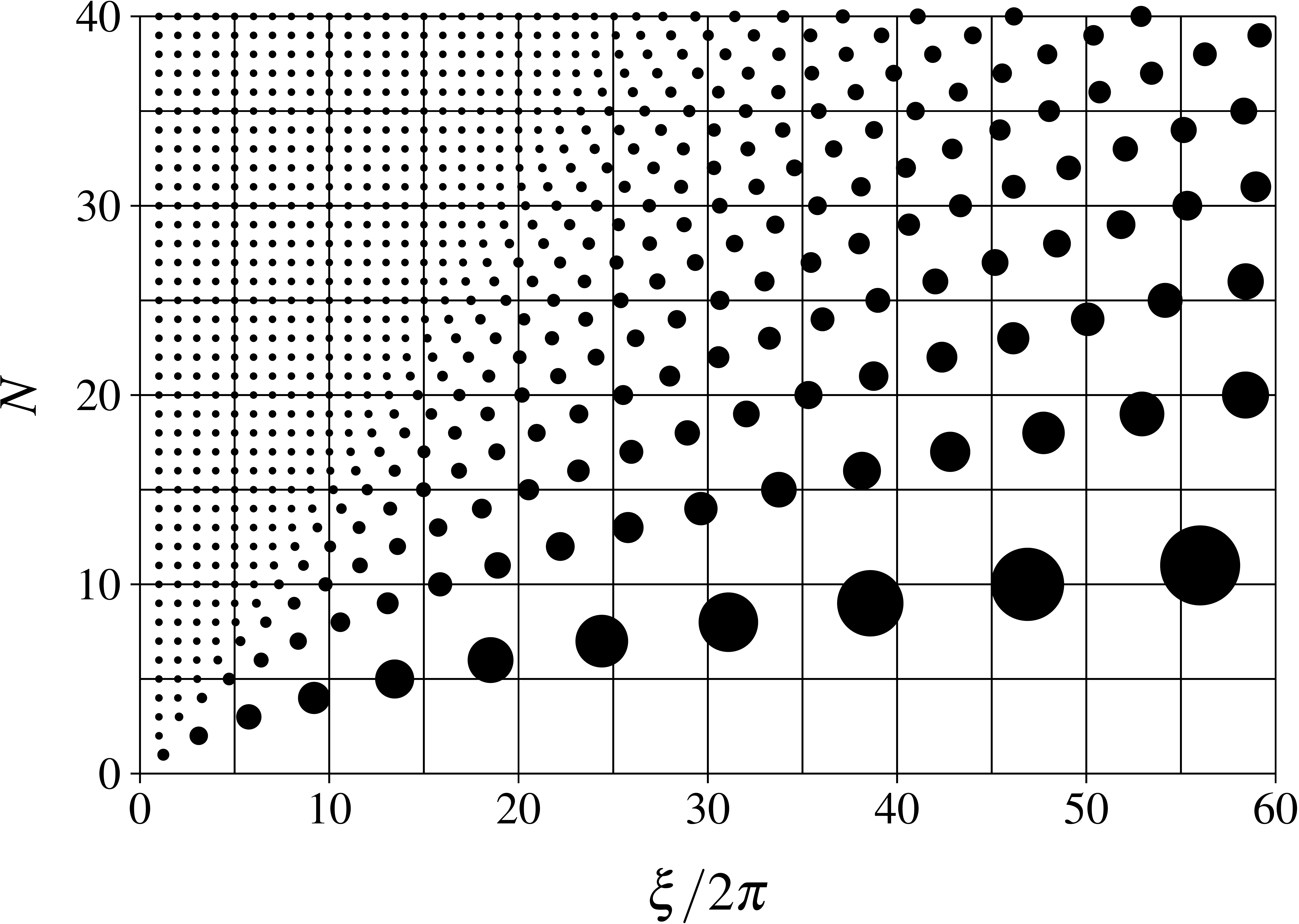}
 \caption{The positive imaginary frequency poles $\xi$ contributing to the
   Pad\'e spectrum decomposition (PSD) are displayed as a function of the order
   $N$ of the Pad\'e approximation. The center of the circle indicates the position
   of the pole while the area of the circle is proportional to the weight associated
   with the pole.}
 \label{fig:psd_poles}
\end{figure}

Figure~\ref{fig:psd_poles} visualizes the poles at imaginary frequencies $\xi>0$
appearing in \eqref{eq:PSD} as a function of the order $N$ of the Pad{\'e}
approximation.  The center of each circle indicates the position of a pole
while the area is proportional to the weight of the pole. For sufficiently
small frequencies, one observes the regularly spaced poles leading to the
Matsubara sum \eqref{eq:F}. For larger frequencies, the spacing of the poles
increases as does their weight. The irregular spacing of the poles implies that
the order of the PSD has to be fixed beforehand.

Within the PSD, the van der Waals free energy is expressed as
\begin{equation}
\mathcal{F} = \frac{k_B T}{2}\left[\Phi(0) + 2\sum_{j=1}^N \eta_j \Phi(i \xi_j)\right]\,.
\end{equation}
Careful numerical tests reveal that convergence of the frequency summation is
reached quicker when using PSD, since the approximation order scales only as
$N\sim \sqrt{\lambda_T/L}$.  Thus, the PSD is superior to the MSD when
computing the van der Waals interaction, in particular for experimentally relevant
system parameters.

\end{document}